\documentclass{elsart}

\usepackage{amssymb}

\begin{document}

% CHANGES as compared to the hep-ph/0403094:
% - in Abstract, ', and' replaced by 'as well as', 'based on' replaced by 
%   'on the basis of';
% - corrected misprinted E^{\mathrm{c}} to E^{\mathrm{e}} in Eq.
%   \label{eq2.02};
% - Eq. \label{eq2.22} has a constant \alpha_{c} on the RHS (error in NuPhA);
% - wrong reference to Table~{\ref{t3.5}} replaced by \ref{t3.4} in the 2nd
%   paragraph after Eq. \label{eq3.03};
% - added the omitted '-' sign to MIT parameter Z_{0} value 1.836 in Table 
%   \label{t3.4} (error in NuPhA);
% - deleted redundant 'more' in 'more a simpler' in 3rd paragraph before Eq.
%   \label{eq3.04};
% - corrected misprinted value of $\Omega _{c}^{0*}$ mass 2.282 to 2.782 in 
%   Table \label{t3.6} (error in NuPhA);
% - corrected misprint $g_{A}$ to $g_{A}^{0}$ in the text after Eq. 
%   \label{eq4.05} (error in NuPhA);
% - replaced 'mass split' by 'mass splitting' in the last paragraph of Summary. 

\begin{frontmatter}



\title{Towards the unified description of light and heavy hadrons in the bag model approach}


\author{A.~Bernotas\corauthref{cor}} and
\ead{bernotas@itpa.lt}
\author{V.~\v{S}imonis}
\ead{simonis@itpa.lt}
\corauth[cor]{Corresponding author.}

\address{Vilnius University Research Institute of Theoretical Physics and Astronomy, A.~Go\v{s}tauto 12, 01108 Vilnius, Lithuania}

\begin{abstract}
% Text of abstract
Mass spectra of ground state hadrons containing \textit{u-}, \textit{d-}, \textit{s-}, \textit{c-} quarks as well as some lightest hadrons containing \textit{b-}quarks are calculated on the basis of a slightly modified bag model. The center-of-mass motion corrections are incorporated using a wavepacket projection with Gaussian parametrization of the distribution amplitude. We use running coupling constant and also allow the effective quark mass to be scale-dependent. The impact of these modifications on the hadron mass spectrum is investigated. A comparison of the predicted mass values with the experimental data demonstrates that the modified bag model is sufficiently flexible to provide a satisfactory description of light and heavy hadrons (mesons and baryons) in a single consistent framework.
\end{abstract}

\end{frontmatter}

% main text
\section{Introduction}
%\label{}

Over the last decade a lot of progress has been made in the experimental spectroscopy of heavy hadrons. Accumulation of the high statistics data by various experiments led to the discovery of many new states. Among others, even rather exotic state of two different heavy quarks ($B_{c}$ meson) \cite{01CDF98,02OPAL98} has been observed. In addition, spectroscopy of heavy hadrons serves as an important field to test various QCD-inspired phenomenological models of hadron structure. One of such models is the MIT (Massachusetts Institute of Technology) bag model \cite{03CJJTW74,04CJJT74}. There are several excellent reviews on this subject available \cite{05J75,06HK78,07S79,08DD83,09T84}, where one can find more information concerning basic equations, applications, and further developments of the bag model.

After the first success in describing the static properties of the light hadrons \cite{10DJJK75}, a straightforward application of the bag model to the heavy quark states \cite{11JK76} led to a surprisingly strong disagreement with the experimental data. Early attempts \cite{12HDDT78,13P79} to improve the model were of limited success. Discrepancies seemed to be of qualitative character, so one could conclude that some more radical modifications of the model were necessary. It was soon realized that the bag model was afflicted by the well-known center-of-mass motion (c.m.m.) problem. A part of the hadron energy calculated in the ordinary bag model is spurious and, consequently, the model must be corrected in some fashion. Such correction may lead to the substantial changes in the predicted mass values of the light hadrons \cite{14DJ80,15W81,16LW82}. To the best of our knowledge, at present there is no unambiguous method to deal with this problem. Nevertheless, approximate schemes have been widely used in various bag model calculations \cite{17CHP83,18FPS84,19D81,20TT85,21BM88,22BSRMT84,23RT85,24BSMT84,25S84,26HM89,27HM90}.

For the hadrons containing one heavy quark an elegant way to eliminate the center-of-mass motion has been proposed \cite{28S80,29IDS82} (for further developments see \cite{30WMM85,31OH99}). In that approach the heavy quark occupies the center of the bag and the light quarks move in the colour field set up by this heavy quark. A simple physical picture is an attractive feature of this prescription. However, its applicability is restricted to hydrogen-like systems. Therefore, if we want to have a unified description of the hadrons, we need a more universal tool to deal with the c.m.m. problem. Although there is some controversy on this subject, we have chosen to follow the technique adopted in Refs. \cite{24BSMT84,25S84,26HM89,27HM90}. The essence of this method is to replace the bag state with the wave packet (a superposition of plane-wave states). A similar approach to correct for the c.m.m. was used within the framework of the relativistic potential model \cite{32ERR85,33BDD85,34PL86,35JPT00}.

The aim of this paper is to provide a unified description of the light and the heavy hadrons in the framework of the bag model. Besides the c.m.m. correction we will incorporate two other QCD-inspired improvements of the model: the running (i.e. scale-dependent) effective coupling constant, and the scale-dependent effective quark mass. The influence of these improvements on the hadron mass spectrum will be investigated.

This paper is organized as follows. In the next section a modified bag model is described. Our results on the calculated hadron spectrum are presented in Section~3 along with a discussion of the influence of the modifications upon the bag model predictions. Some other static parameters of light hadrons (magnetic moments, axial-vector coupling constant $g_{A}$, and charge radii) for which experimental data exist are calculated and presented in Section~4. Finally, we summarize our conclusions in Section~5.

\section{The model}

The bag model enables us to calculate the static properties of hadrons by making a number of simplifying assumptions. Usually it is assumed that the quarks are confined in the sphere of fixed radius $R$, within which they obey the free Dirac equation (static spherical cavity approximation). The energy of a hadron is given by 
\begin{equation}
E=\frac{4\pi }{3}BR^{3}+\sum\limits_{i}n_{i}\varepsilon _{i}+\Delta E.
\label{eq2.01}
\end{equation}

The first term on the right-hand side of the above equation is the bag volume energy that guarantees the quark confinement in the finite region, $R$ stands for the radius of the bag, and $B$ is the bag constant. The second term is the ``kinetic'' energy of quarks, $n_{i}$ is the number of quarks of \textit{i}-th flavour, $\varepsilon _{i}$ -- the eigenenergy of a quark in the cavity. The last term represents the interaction energy of the quarks in the Abelian approximation to QCD. Minimization of the energy determines the bag radius $R_{0}$ of the hadron under consideration.

It is useful to divide $\Delta E$ into two parts:
\begin{equation}
\Delta E=E^{\mathrm{m}}+E^{\mathrm{e}}.  \label{eq2.02}
\end{equation}
One,
\begin{equation}
E^{\mathrm{m}}=\alpha _{c}\left[
\sum\limits_{i}a_{ii}M_{ii}+\sum\limits_{j>i}a_{ij}M_{ij}\right] ,
\label{eq2.03}
\end{equation}
is the colour-magnetic part, and another,
\begin{equation}
E^{\mathrm{e}}=\alpha _{c}\left[
\sum\limits_{i}f_{i}I_{ii}+\sum\limits_{j>i}f_{ij}I_{ij}\right] ,
\label{eq2.04}
\end{equation}
is the colour-electric (Coulomb) part of the interaction. In Eqs.~(\ref{eq2.03}) and (\ref{eq2.04}) $\alpha _{c}$ is the coupling constant and the sum runs over the flavour indices. For the benefit of the reader, below we present the expressions (\ref{eq2.03}) and (\ref{eq2.04}) in more detail, omitting tedious derivation procedures. Functions $M_{ij}(R)$ and $I_{ij}(R)$ can be written in the form
\begin{equation}
M_{ij}(R)=\frac{4}{3}\int\limits_{0}^{R}dr\mu _{i}^{\prime }(r)A_{j}(r,R),
\label{eq2.05}
\end{equation}

\begin{equation}
I_{ij}(R)=\frac{2}{3}\int\limits_{0}^{R}dr\rho _{i}^{\prime }(r)V_{j}(r,R).
\label{eq2.06}
\end{equation}
Here
\begin{equation}
\mu _{i}^{\prime }(r)=-\frac{2r}{3}P_{i}(r)Q_{i}(r)  \label{eq2.07}
\end{equation}
is the scalar magnetization density of an \textit{i}-th quark. The semiclassical vector potential generated by the \textit{i}-th quark has the form \cite{36GHK83}
\begin{equation}
A_{i}(r,R)=\frac{\mu _{i}(r)}{r^{3}}+\frac{\mu _{i}(R)}{2R^{3}}+\mathrm{M}%
_{i}(r,R),  \label{eq2.08}
\end{equation}
where
\begin{equation}
\mu _{i}(r)=\int\limits_{0}^{r}dx\mu _{i}^{\prime }(x),  \label{eq2.09}
\end{equation}

\begin{equation}
\mathrm{M}_{i}(r,R)=\int\limits_{r}^{R}dx\frac{\mu _{i}^{\prime }(x)}{x^{3}}.
\label{eq2.10}
\end{equation}
In Eq.~(\ref{eq2.06})
\begin{equation}
\rho _{i}^{\prime }(r)=P_{i}^{2}\left( r\right) +Q_{i}^{2}\left( r\right)
\label{eq2.11}
\end{equation}
is the charge density of the \textit{i}-th quark, and
\begin{equation}
\rho _{i}(r)=\int\limits_{0}^{r}dx\rho _{i}^{\prime }(x),  \label{eq2.12}
\end{equation}

\begin{equation}
V_{i}(r,R)=\rho _{i}(r)\left( \frac{1}{r}-\frac{1}{R}\right)
+\int\limits_{r}^{R}dx\frac{\rho _{i}^{\prime }(x)}{x}.  \label{eq2.13}
\end{equation}
$P_{i}(r)$ and $Q_{i}(r)$ in Eqs.~(\ref{eq2.07}) and (\ref{eq2.11}) are the large and small radial functions of the two-component spherical spinor normalized as
\begin{equation}
\int\limits_{0}^{R}dr\left[ P_{i}^{2}\left( r\right) +Q_{i}^{2}\left(
r\right) \right] =1,  \label{eq2.14}
\end{equation}
and obeying the linear boundary condition
\begin{equation}
P_{i}(R)=-Q_{i}(R)  \label{eq2.15}
\end{equation}
at the bag surface.

In order to avoid any possible complications we have used the confined Coulomb Green's function \cite{37L79} in the derivation of the expression~(\ref {eq2.06}). As a consequence, the value of the colour scalar potential $V_{i}(r,R)$ is zero at the surface of the cavity:
\begin{equation}
V_{i}(R,R)=0 \,.  \label{eq2.16}
\end{equation}
The coefficients $a_{ij}$, $f_{i}$, and $f_{ij}$ that specify the interaction energy of hadrons in Eqs.~(\ref{eq2.03}) and (\ref{eq2.04}) can be readily calculated using the technique described in Ref.~\cite{10DJJK75}. Parameters $f_{i}$ that specify the colour-electrostatic interaction energy between the quarks of the same flavour are
\begin{equation}
f_{i}=-\lambda \cdot n_{i}(n_{i}-1)/2,  \label{eq2.17}
\end{equation}
and parameters $f_{ij}$ ($i\neq j$) are given by
\begin{equation}
f_{ij}=-\lambda \cdot n_{i}n_{j},  \label{eq2.18}
\end{equation}
where
\begin{equation}
\lambda =\left\{ 
\begin{array}{rl}
1 & \quad \textrm{for baryons,} \\ 
2 & \quad \textrm{for mesons.}
\end{array}
\right.  \label{eq2.19}
\end{equation}

Parameters $a_{ij}$ that specify the colour-magnetostatic interaction energy for mesons with the total spin $J$ are
\begin{equation}
a_{ij}=\left\{ 
\begin{array}{rl}
-6 & \quad (J=0), \\ 
2 & \quad (J=1).
\end{array}
\right.  \label{eq2.20}
\end{equation}
For the baryons consisting of\textit{\ u-, d-, s-, }and\textit{\ c-}quarks these parameters are given in Table~\ref{t2.1}. For the baryons containing \textit{b-}quarks the corresponding parameters can be easily defined by means of simple substitutions (e.g., $c\rightarrow b$). In the case of the light hadrons one can also find the parameters $a_{ij}$ in Table~2 of Ref.~\cite{10DJJK75}.
%TCIMACRO{\TeXButton{B}{\begin{table}[tbp] \centering}}
%BeginExpansion
\begin{table}[tbp] \centering%t01
%EndExpansion
\caption{Parameters that specify the colour-magnetic interaction energy of
baryons consisting of the light $(l=u,d)$, strange $(s)$, and charmed $(c)$
quarks.\label{t2.1}}
\begin{tabular}{|c|c|c|c|c|c|c|c|c|}
\hline
$J$ & Particle & Quark content & $a_{ll}$ & $a_{ls}$ & $a_{lc}$ & $a_{ss}$ & 
$a_{sc}$ & $a_{cc}$ \\ \hline
$1/2$ & $N$ & $lll$ & $-3$ &  &  &  &  &  \\ \hline
$1/2$ & $\Lambda $ & $s(ll)_{anti}$ & $-3$ &  &  &  &  &  \\ \hline
$1/2$ & $\Sigma $ & $s(ll)_{sym}$ & $1$ & $-4$ &  &  &  &  \\ \hline
$1/2$ & $\Xi $ & $lss$ &  & $-4$ &  & $1$ &  &  \\ \hline
$1/2$ & $\Lambda _{c}^{+}$ & $c(ll)_{anti}$ & $-3$ &  &  &  &  &  \\ \hline
$1/2$ & $\Sigma _{c}$ & $c(ll)_{sym}$ & $1$ &  & $-4$ &  &  &  \\ 
\hline
$1/2$ & $\Xi _{c}$ & $c(ls)_{anti}$ &  & $-3$ &  &  &  &  \\ \hline
$1/2$ & $\Xi _{c}^{\prime }$ & $c(ls)_{sym}$ &  & $1$ & $-2$ &  & $-2$ &  \\ 
\hline
$1/2$ & $\Omega _{c}^{0}$ & $css$ &  &  &  & $1$ & $-4$ &  \\ \hline
$1/2$ & $\Xi _{cc}$ & $lcc$ &  &  & $-4$ &  &  & $1$ \\ \hline
$1/2$ & $\Omega _{cc}^{+}$ & $scc$ &  &  &  &  & $-4$ & $1$ \\ \hline
$3/2$ & $\Delta $ & $lll$ & $3$ &  &  &  &  &  \\ \hline
$3/2$ & $\Sigma ^{*}$ & $sll$ & $1$ & $2$ &  &  &  &  \\ \hline
$3/2$ & $\Xi ^{*}$ & $lss$ &  & $2$ &  & $1$ &  &  \\ \hline
$3/2$ & $\Omega ^{-}$ & $sss$ &  &  &  & $3$ &  &  \\ \hline
$3/2$ & $\Sigma _{c}^{*}$ & $cll$ & $1$ &  & $2$ &  &  &  \\ \hline
$3/2$ & $\Xi _{c}^{*}$ & $cls$ &  & $1$ & $1$ &  & $1$ &  \\ \hline
$3/2$ & $\Omega _{c}^{0*}$ & $css$ &  &  &  & $1$ & $2$ &  \\ \hline
$3/2$ & $\Xi _{cc}^{*}$ & $lcc$ &  &  & $2$ &  &  & $1$ \\ \hline
$3/2$ & $\Omega _{cc}^{+*}$ & $scc$ &  &  &  &  & $2$ & $1$ \\ \hline
$3/2$ & $\Omega _{ccc}^{++}$ & $ccc$ &  &  &  &  &  & $3$ \\ \hline
\end{tabular}
%TCIMACRO{\TeXButton{E}{\end{table}}}
%BeginExpansion
\end{table}%
%EndExpansion

Now we describe the salient differences between our treatment and the original version of the MIT bag model. In the expression of the bag energy~(\ref{eq2.01}) we have omitted two terms that were present in the MIT version \cite{10DJJK75,11JK76} of the model,
\begin{equation}
E_{0}=\frac{Z_{0}}{R},  \label{eq2.21}
\end{equation}
and
\begin{equation}
E_{\mathrm{self}}^{\mathrm{e}}=\alpha_{c}\sum\limits_{i}n_{i}I_{ii}.  \label{eq2.22}
\end{equation}
The first term was expected to represent the so-called zero-point (Casimir) energy. This entry was necessary to obtain a good fit in the original MIT version of the bag model. However, the phenomenological value of the parameter $Z_{0}\simeq -1.9$ differs substantially from its theoretical value $Z_{0}\simeq +0.7$ \cite{38M83}. Note that even the sign of the effect is opposite. As shown in \cite{16LW82}, the phenomenological value of $Z_{0}$ can be made smaller by introducing the c.m.m. correction and refitting model parameters.

The second term represents a part of the self-energy of quarks included in a somewhat arbitrary fashion. Such a choice reduces substantially the colour-electric part of the interaction energy and in the case of quarks of the same mass makes it vanish. This is to be contrasted to the potential model in which the Coulomb-like colour-electric potential plays an essential role in the description of the $J/\psi $ and $\eta _{c}$ mesons (see also Refs. \cite{39SV98} and \cite{06HK78} for the critical discussion on this subject). We think that the description of the same states in two models must not be so different. Furthermore, in the usual approach all the self-energy can be absorbed into the renormalization of the quark mass and, therefore, any use of the self-energy term in the energy expression could cause a double counting. So, in order to have a consistent description of the heavy hadrons \cite{39SV98} we have chosen to discard the term~(\ref{eq2.22}) from the bag model energy.

Now let us proceed with the further modifications we want to include in our version of the bag model. First of all, QCD guidelines should be followed where possible. We will incorporate two QCD inspired modifications: scale dependence of the strong coupling constant $\alpha _{c}$ and scale dependence of the effective quark mass $m_{f}$. Determination of the quark mass is a very interesting and complicated problem by itself. Due to the confinement, quarks are not the asymptotic states of QCD and, therefore, their masses cannot be measured directly. Moreover, the mass values depend on the chosen conventions and can be determined only through their influence on the properties of hadrons. The problem of the determination of quark masses in the context of the heavy quark theory is discussed in \cite{40BSU97}, and some properties of the effective quark mass are studied in \cite{41ES88}. For the recent review see the article by A.V.~Manohar in \cite{42PDG02}.

Strictly speaking, there is no way to relate the quark mass as defined in the phenomenological models (such as potential model or bag model) to the parameters of the QCD Lagrangian, or to the pole mass. Many ingredients of the phenomenological models are introduced by hand and can be justified only by the success of the model in describing the experimentally measured properties of the hadrons. Nevertheless, we expect these models to share some of their features with QCD. What can QCD tell us about the properties of the strong coupling constant and quark mass? From the renormalization-group analysis, with the $n_{f}$ quark flavours for which $m_{f}<<Q$, in the leading logarithmic approximation there follows \cite{43Y83}:
\begin{equation}
\alpha _{s}(Q^{2})=\frac{12\pi }{(33-2n_{f})\ln (Q^{2}/\Lambda ^{2})},
\label{eq2.23}
\end{equation}

\begin{equation}
\overline{m}(Q^{2})=\frac{\widehat{m}}{\left[ \frac{1}{2}\ln (Q^{2}/\Lambda
^{2})\right] ^{d_{m}}},  \label{eq2.24}
\end{equation}
where $\overline{m}(Q^{2})$ is the mass function (running mass) in the $\overline{\mathrm{MS}}$ scheme, $\Lambda \simeq 200$~MeV -- the QCD constant, $\widehat{m}$ -- some new integration constant (analogue of $\Lambda $), and $d_{m}=12/(33-2n_{f})$ -- anomalous dimension of the mass.

We are working in the soft regime where the behaviour of Eqs.~(\ref{eq2.23}) and (\ref{eq2.24}) is not well-defined. So, instead of Eq.~(\ref{eq2.23}) we will employ the cavity-radius-dependent parametrization proposed in Ref.~\cite{14DJ80}:
\begin{equation}
\alpha _{s}(R)=\frac{2\pi }{9\ln (A+R_{0}/R)}  \label{eq2.25}
\end{equation}
consistent with Eq.~(\ref{eq2.23}). Parameter $A$ helps us to avoid divergences when $R\rightarrow R_{0}$, where $R_{0}$ is the scale parameter analogous to QCD constant $\Lambda $ in the momentum space. An alternative choice could be the \textit{r}-dependent function obtained using the procedure adopted in \cite{44GI85,45CI86} in the context of the relativized potential model. For the time being we prefer to use the expression~(\ref{eq2.25}) because of its simplicity. We expect it to provide some average estimate of the scale dependence of the strength of the effective interaction inside the bag.

 For the effective quark mass we employ the parametrization
\begin{equation}
\overline{m}_{f}(R)=\widetilde{m}_{f}+\alpha _{s}(R)\cdot \delta _{f},  \label{eq2.26}
\end{equation}
with two flavour-dependent parameters $\widetilde{m}_{f}$ and $\delta _{f}$. Despite rather different form, there is no serious contradiction between Eqs.~(\ref{eq2.26}) and (\ref{eq2.24}). In the sufficiently wide range of parameters Eq.~(\ref{eq2.26}) can be approximated by
\begin{equation}
\overline{m}_{f}(R)=\frac{\widehat{m}_{f}}{\left[ \frac{1}{2}\ln
(C_{f}+R_{0}/R)\right] ^{d_{m}}}  \label{eq2.27}
\end{equation}
with two other flavour-dependent parameters $\widehat{m}_{f}$ and $C_{f}$. Equation~(\ref{eq2.27}) can be interpreted as divergency-free extension of Eq.~(\ref{eq2.24}).

Scale dependence of the quark mass proved to be important in the relativistic flux tube model calculations \cite{46OV95}. Asymptotic behaviour of the mass function in the framework of the Bethe--Salpeter equation coupled to the Schwinger--Dyson equation for the quark propagators \cite{47MJ92} also agrees well with Eq.~(\ref{eq2.24}).

 It is hard to obtain equally good description of mesons and baryons in the quark model with the common value for the quark mass. In order to improve the description K.~Cahill \cite{48C91} proposed to use two sets of constituent quark masses: one set for the constituents of mesons, and another set for the constituents of baryons. Since the bag model also suffers from the flaw of this kind, we expect that the introduction of the scale-dependent effective mass would help us to improve the situation in this case as well.

To proceed with the calculations of the hadronic properties we must relate the ground state energy~(\ref{eq2.01}) to the mass of the hadron. To this end we adopt the procedure proposed in \cite{24BSMT84,25S84} and consider the bag state $\left| B\right\rangle $ as a wave packet of the physical states $\left| B,\mathbf{p}\right\rangle $ with various total momenta 
\begin{equation}
\left| B\right\rangle =\int d^{3}p\Phi (|\mathbf{p}|)\left| B,\mathbf{p}%
\right\rangle .  \label{eq2.28}
\end{equation}

In general \cite{49L58}, equation of this type cannot be exact (non-relativistic harmonic oscillator being an exception). So, we do not expect Eq.~(\ref{eq2.28}) to provide the exact solution to the c.m.m. problem and consider this relation as a reasonable ansatz only. For the profile function we adopt a Gaussian parametrization \cite{26HM89,27HM90,32ERR85}:
\begin{equation}
\Phi _{P}(s)=\left( \frac{3}{2\pi P^{2}}\right) ^{\frac34 } \exp \left( -\frac{3s^{2}}{4P^{2}} \right) .  \label{eq2.29}
\end{equation}
Functions $\Phi _{P}(s)$ are normalized as 
\begin{equation}
\int d^{3}s\Phi _{P}^{2}(s)=1.  \label{eq2.30}
\end{equation}
Parameter $P$ that specifies the momentum distribution still needs to be determined by making some reasonable assumption. We will use the prescription
\begin{equation}
P^{2}=\gamma \sum\limits_{i}n_{i}p_{i}^{2},  \label{eq2.31}
\end{equation}
where $p_{i}=\left( \varepsilon _{i}^{2}-m_{i}^{2}\right) ^{\frac{1}{2}}$ is the momentum of the\textit{\ i}-th quark. The c.m.m. parameter $\gamma$ will be determined in the fitting procedure. At the first sight, the natural choice seems to be $\gamma = 1$, however, for the reasons that will be discussed later we will use a more general form~(\ref{eq2.31}) and interpret $P^{2}$ as an effective momentum square.

All averages have to be calculated with the profile $\Phi _{P}(s)$. In the following we will need the quantities $\left\langle E\right\rangle $, $\left\langle M/E\right\rangle $, and $\left\langle M^{2}/E^{2}\right\rangle$:
\begin{equation}
\left\langle E\right\rangle =\int d^{3}s\Phi _{P}^{2}(s)\sqrt{M^{2}+s^{2}},
\label{eq2.32}
\end{equation}

\begin{equation}
\left\langle \frac{M}{E}\right\rangle =\int d^{3}s\Phi _{P}^{2}(s)\frac{M}{ \sqrt{M^{2}+s^{2}}},  \label{eq2.33}
\end{equation}

\begin{equation}
\left\langle \frac{M^{2}}{E^{2}}\right\rangle =\int d^{3}s\Phi _{P}^{2}(s) \frac{M^{2}}{M^{2}+s^{2}}.  \label{eq2.34}
\end{equation}

By using (\ref{eq2.29}), Eq.~(\ref{eq2.32}) can be rewritten as 
\begin{equation}
\left\langle E\right\rangle =\sqrt{\frac{54}{\pi }}\int\limits_{0}^{\infty}s^{2}ds\sqrt{P^{2}s^{2}+M^{2}} \, \exp \left( -\frac{3}{2}s^{2} \right) .
\label{eq2.35}
\end{equation}

Once the energy of an individual hadron $E$ and the effective momentum $P$ are given, Eq.~(\ref{eq2.35}) can be solved to obtain the mass of the particle (see Ref.~\cite{20TT85} for somewhat different procedure).

As noted in \cite{27HM90}, from Eq.~(\ref{eq2.35}) one can easily obtain the relation
\begin{equation}
M^{2}=\left\langle E\right\rangle ^{2}-\beta \left( \frac{M^{2}}{P^{2}}%
\right) P^{2},  \label{eq2.36}
\end{equation}
where
\begin{equation}
\beta (x)=\frac{54}{\pi }\left( \int\limits_{0}^{\infty }t^{2}dt\sqrt{t^{2}+x} \, \exp \left(-\frac{3}{2}t^{2}\right)\right) ^{2}-x.  \label{eq2.37}
\end{equation}
The limiting values of this function 0.85 and 1 correspond to ultra-relativistic and non-relativistic cases, respectively. Equation~(\ref{eq2.36}) looks much like the familiar Einstein relation
\begin{equation}
M^{2}=E^{2}-P^{2},  \label{eq2.38}
\end{equation}
that is very popular in the various bag-model-based calculations.

\section{Results. Hadron mass spectrum}

In this section we present the calculated mass values of the ground state hadrons (Tables~\ref{t3.1}--\ref{t3.3}) and analyze the influence of several modifications on the predictions of the model. We begin with the traditional version of the MIT bag model \cite{10DJJK75,11JK76}. The standard expression for the mass of the hadron in this model can be written as 
\begin{equation}
M_{\mathrm{MIT}}=E+E_{\mathrm{self}}^{\mathrm{e}}+E_{0},  \label{eq3.01}
\end{equation}
where the entries in the right-hand side are given by Eqs.~(\ref{eq2.01}), (\ref{eq2.21}), and (\ref{eq2.22}). We adopt the same model parameters $B$, $Z_{0}$, $\alpha _{c}$, $m_{s}$, $m_{c}$ (see Table~\ref{t3.4}) as in the original treatment \cite{10DJJK75,11JK76} and use the experimental mass value of the $\Upsilon$ meson to determine the mass of the \textit{b}-quark $m_{b}$. The up and down quarks are taken to be massless.
%TCIMACRO{\TeXButton{B}{\begin{table}[tbp] \centering}}
%BeginExpansion
\begin{table}[tbp] \centering%t02
%EndExpansion
\caption{Mases (in GeV) of hadrons consisting of the \textit{u-, d-}, and \textit{s-}quarks in the six variants of the bag model as described in the text. Underlined entries were used to define the free model parameters. $\chi_{_{12}}$ was calculated without the contribution of the pseudoscalar mesons.\label{t3.1}}
\begin{tabular}{llllllll}
\hline
Particle & $M_{\mathrm{ex}}$ & $M_{\mathrm{MIT}}$ & Mod1 & Mod2 & Mod3 & Mod4 & Mod5 \\ \hline
$N$ & 0.939 & \underline{0.938} & \underline{0.939} & \underline{0.939} & \underline{0.939} & \underline{0.939} & \underline{0.939} \\ 
$\Delta $ & 1.232 & \underline{1.233} & \underline{1.232} & \underline{1.232} & \underline{1.232} & \underline{1.232} & \underline{1.232} \\ 
$\pi $ & 0.137 & 0.280 & -- & -- & \underline{0.137} & \underline{0.137} & 0.252 \\ 
$\rho $ & 0.769 & 0.783 & 0.776 & 0.776 & 0.776 & 0.776 & 0.776 \\ 
$\omega $ & 0.782 & \underline{0.783} & \underline{0.776} & \underline{0.776} & \underline{0.776} & \underline{0.776} & \underline{0.776} \\ 
$\Lambda $ & 1.116 & 1.104 & 1.099 & 1.101 & 1.098 & \underline{1.116} & \underline{1.116} \\ 
$\Sigma $ & 1.193 & 1.144 & 1.140 & 1.143 & 1.138 & 1.159 & 1.158 \\ 
$\Xi $ & 1.318 & 1.288 & 1.280 & 1.283 & 1.277 & 1.310 & 1.310 \\ 
$\Sigma ^{*}$ & 1.385 & 1.382 & 1.372 & 1.374 & 1.368 & 1.388 & 1.384 \\ 
$\Xi ^{*}$ & 1.533 & 1.528 & 1.513 & 1.517 & 1.505 & 1.543 & 1.536 \\ 
$\Omega ^{-}$ & 1.672 & \underline{1.672} & 1.654 & 1.660 & 1.643 & 1.695 & 1.687 \\ 
$K$ & 0.496 & 0.496 & 0.326 & 0.324 & 0.458 & 0.437 & \underline{0.496} \\ 
$K^{*}$ & 0.894 & 0.928 & 0.896 & 0.896 & 0.895 & 0.897 & 0.895 \\ 
$\phi $ & 1.019 & 1.067 & \underline{1.019} & \underline{1.019} & \underline{1.019} & \underline{1.019} & \underline{1.019} \\ 
$\chi_{_{12}}$ & -- & 0.023 & 0.021 & 0.019 & 0.024 & 0.013 & 0.012 \\ \hline
\end{tabular}
%TCIMACRO{\TeXButton{E}{\end{table}}}
%BeginExpansion
\end{table}%
%EndExpansion

The empirical zero-point energy term $E_{0}$ used in the original version of the MIT bag model was later reinterpreted as representing mostly a c.m.m. correction \cite{16LW82}. As the first step in modifying the model we omit this term and use a more elaborated procedure based on Eq.~(\ref{eq2.35}) to account for the c.m.m. In this variant of the model (denoted as Mod1) the energy is given by
\begin{equation}
E_{\mathrm{Mod1}}=E+E_{\mathrm{self}}^{\mathrm{e}},  \label{eq3.02}
\end{equation}
which is minimized in order to determine the radius $R$ of the spherical cavity in which the hadron is confined. After the minimization is performed, Eq.~(\ref{eq2.35}) must be solved numerically to obtain the mass $M$ of the corresponding hadron. The free parameters of the model now are $B$, $\gamma$, $\alpha _{c}$, $m_{s}$, $m_{c}$, and $m_{b}$. Instead of $Z_{0}$ now we have another free parameter $\gamma$ governing the c.m.m. correction. To fix $B$, $\gamma$, and $\alpha _{c}$, the masses of the light baryons $N$, $\Delta $, and the average mass of the $\omega $--$\rho $ system are employed. We use the vector meson $\phi $ (instead of the baryon $\Omega ^{-}$ used in the MIT version of the model) to fix the strange quark mass. Our choice is motivated by an intent to have the same mass fixing procedure for \textit{s-, c-}, and \textit{b-}quarks. For this purpose we will employ the mass of the corresponding vector meson (i.e. $\phi $, $J/\psi $, and $\Upsilon $).

The experimental mass values of the hadrons were taken from the Particle Data Group \cite{42PDG02}. For the isospin multiplets the averaged values were used.
%TCIMACRO{\TeXButton{B}{\begin{table}[tbp] \centering}}
%BeginExpansion
\begin{table}[tbp] \centering%t03
%EndExpansion
\caption{Mases (in GeV) of hadrons containing charmed quarks in the six variants of the bag model as described in the text. Underlined entries were used to define the free model parameters.\label{t3.2}} 
\begin{tabular}{llllllll}
\hline
Particle & $M_{\mathrm{ex}}$ & $M_{\mathrm{MIT}}$ & Mod1 & Mod2 & Mod3 & Mod4 & Mod5 \\ \hline
$\Lambda _{c}^{+}$ & 2.285 & 2.215 & 2.259 & 2.279 & 2.257 & \underline{2.285} & \underline{2.285} \\ 
$\Sigma _{c}$ & 2.452 & 2.358 & 2.373 & 2.393 & 2.364 & 2.392 & 2.389 \\ 
$\Xi _{c}$ & 2.469 & 2.397 & 2.431 & 2.451 & 2.429 & 2.466 & 2.467 \\ 
$\Xi _{c}^{\prime }$ & 2.576 & 2.508 & 2.518 & 2.538 & 2.508 & 2.546 & 2.543 \\ 
$\Omega _{c}^{0}$ & 2.698 & 2.654 & 2.662 & 2.683 & 2.652 & 2.696 & 2.694 \\ 
$\Xi _{cc}$ & -- & 3.540 & 3.527 & 3.552 & 3.520 & 3.556 & 3.554 \\ 
$\Omega _{cc}^{+}$ & -- & 3.691 & 3.677 & 3.702 & 3.671 & 3.709 & 3.709 \\ 
$\Sigma _{c}^{*}$ & 2.518 & 2.462 & 2.464 & 2.491 & 2.458 & 2.488 & 2.482 \\ 
$\Xi _{c}^{*}$ & 2.646 & 2.604 & 2.602 & 2.630 & 2.594 & 2.637 & 2.629 \\ 
$\Omega _{c}^{0*}$ & -- & 2.743 & 2.740 & 2.768 & 2.729 & 2.782 & 2.774 \\ 
$\Xi _{cc}^{*}$ & -- & 3.663 & 3.628 & 3.668 & 3.616 & 3.659 & 3.649 \\ 
$\Omega _{cc}^{+*}$ & -- & 3.797 & 3.763 & 3.801 & 3.751 & 3.799 & 3.791 \\ 
$\Omega _{ccc}^{++}$ & -- & 4.830 & 4.751 & 4.784 & 4.738 & 4.776 & 4.769 \\ 
$D$ & 1.867 & 1.726 & 1.806 & 1.796 & 1.830 & 1.833 & 1.849 \\ 
$D^{*}$ & 2.008 & 1.970 & 1.994 & 2.007 & 1.990 & 2.002 & 1.998 \\ 
$D_{s}$ & 1.969 & 1.886 & 1.947 & 1.936 & 1.975 & 1.965 & 1.986 \\ 
$D_{s}^{*}$ & 2.112 & 2.100 & 2.113 & 2.124 & 2.112 & 2.119 & 2.118 \\ 
$\eta _{c}$ & 2.980 & 2.933 & 2.999 & 2.964 & 3.018 & 3.005 & 3.020 \\ 
$J/\psi $ & 3.097 & \underline{3.097} & \underline{3.097} & \underline{3.097} & \underline{3.097} & \underline{3.097} & \underline{3.097} \\ 
$\chi_{_{13}}$ & -- & 0.071 & 0.042 & 0.032 & 0.046 & 0.024 & 0.027 \\ \hline
\end{tabular}
%TCIMACRO{\TeXButton{E}{\end{table}}}
%BeginExpansion
\end{table}%
%EndExpansion

It is difficult to assess the efficacy of the different variants of the model simply by examining the columns of numbers presented. To assist the reader, in the last row of the Tables~\ref{t3.1}--\ref{t3.3} the $\chi_{_{N}}$ values for each of the variants are presented. This quantity -- a root mean squared deviation from the experimental mass spectra -- is evaluated as follows:
\begin{equation}
\chi_{_{N}}=\left[ \frac{1}{N}\sum\limits_{i=1}^{N}\left(
M^{i}-M_{\mathrm{ex}}^{i}\right) ^{2}\right] ^{\frac{1}{2}},  \label{eq3.03}
\end{equation}
where $M^{i}$ is the model prediction for the \textit{i}-th hadron, $M_{\mathrm{ex}}^{i}$ is the experimental value, and the sum includes $N$ states for which sufficiently accurate values of $M_{\mathrm{ex}}$ are available.
%TCIMACRO{\TeXButton{B}{\begin{table}[tbp] \centering}}
%BeginExpansion
\begin{table}[tbp] \centering%t04
%EndExpansion
\caption{Mases (in GeV) of the lightest hadrons containing bottom quarks in the six variants of the bag model as described in the text. Underlined entries were used to define free model parameters. The masses of $B_{c}$ and $\eta _{b}$ were not included when determining $\chi_{_{6}}$. $M_{\mathrm{ex}}$ of $B_{c}$ was taken from \cite{02OPAL98}. \label{t3.3}}
\begin{tabular}{llllllll}
\hline
Particle & $M_{\mathrm{ex}}$ & $M_{\mathrm{MIT}}$ & Mod1 & Mod2 & Mod3 & Mod4 & Mod5 \\ \hline
$\Lambda _{b}^{0}$ & 5.624 & 5.548 & 5.580 & 5.695 & 5.593 & \underline{5.624} & \underline{5.624} \\ 
$\Sigma _{b}$ & -- & 5.746 & 5.719 & 5.839 & 5.728 & 5.759 & 5.756 \\ 
$B$ & 5.279 & 5.148 & 5.226 & 5.304 & 5.235 & 5.252 & 5.255 \\ 
$B^{*}$ & 5.325 & 5.253 & 5.283 & 5.378 & 5.290 & 5.309 & 5.306 \\ 
$B_{s}$ & 5.370 & 5.283 & 5.361 & 5.435 & 5.374 & 5.387 & 5.393 \\ 
$B_{s}^{*}$ & 5.417 & 5.379 & 5.412 & 5.504 & 5.422 & 5.439 & 5.439 \\ 
$B_{c}$ & 6.32$\pm$0.06 & 6.217 & 6.297 & 6.315 & 6.307 & 6.307 & 6.314 \\ 
$B_{c}^{*}$ & -- & 6.331 & 6.335 & 6.385 & 6.340 & 6.345 & 6.345 \\ 
$\eta _{b}$ & 9.30$\pm$0.04 & 9.258 & 9.438 & 9.374 & 9.441 & 9.438 & 9.442 \\ 
$\Upsilon $ & 9.460 & \underline{9.460} & \underline{9.460} & \underline{9.460} & \underline{9.460} & \underline{9.460} & \underline{9.460} \\ 
$\chi_{_{6}}$ & -- & 0.083 & 0.033 & 0.058 & 0.026 & 0.017 & 0.018 \\ \hline
\end{tabular}
%TCIMACRO{\TeXButton{E}{\end{table}}}
%BeginExpansion
\end{table}%
%EndExpansion

By comparing the results of our calculation (presented in the column denoted as Mod1 in Tables~\ref{t3.1}--\ref{t3.3}) with the predictions of the original version of the MIT bag model and with the experimental values we see that for the hadrons containing heavy quarks (Tables~\ref{t3.2} and \ref{t3.3}) the agreement between predicted and experimental values is obviously improved. For the light hadrons the agreement with experiment is of the quality similar to the MIT version. The pseudoscalar mesons ($\pi $ and $K$) remain the source of some difficulty. The kaon comes out about 170~MeV too low, and there is no solution for the pion. The small masses of these mesons result after the partial cancellation of several large terms. As a consequence, these mass values are strongly model-dependent and rather sensitive to the changes of the model parameters.

Table~\ref{t3.4} shows that the bag constant $B$ and the mass parameters $m_{f}$ have not changed substantially from the original MIT version. The strong coupling constant $\alpha _{c}$ has reduced from 2.19 to 1.56. It can be noted also that for the light hadrons our predictions are qualitatively similar to the results obtained in Refs.~\cite{16LW82,17CHP83,23RT85}.

Now, let us make the next step and drop out the self-energy term. The new version of the model (denoted as Mod2) coincides with the preceding one with the only exception that energy of the hadron is now given by Eq.~(\ref{eq2.01}) instead of (\ref{eq3.02}). There are practically no changes in the predictions of the light hadron masses (only the mass of $\Omega ^{-}$ is slightly improved), while the model parameters undergo sizeable changes. This can be considered as some kind of renormalization, since the effect of the self-energy term now must be absorbed in the redefinition of the parameters of the model. In the heavy quark sector the predictions of the two versions (with and without the self-energy term) differ. For the hadrons with charm the overall fit is improved again. For the hadrons containing bottom quarks the situation is opposite. The agreement with experiment in the new variant is somewhat spoiled. The fit is still better than in the case of the original MIT version, however, it can hardly be considered as satisfactory.

Now we are in the position to examine the influence of the scale dependence of the effective strong coupling constant on the predictions of the bag model, by replacing $\alpha _{c}$ with Eq.~(\ref{eq2.25}). In this new variant of the model we have one extra parameter (say, $A$), and therefore, in order to determine its value, some extra prescription is necessary. Following Ref.~\cite{29IDS82} one can put $A=1$ in Eq.~(\ref{eq2.25}), and this would not be a bad choice. Another possible choice could be the requirement for the pion mass to vanish when $m_{\mathrm{q}}\rightarrow 0$ \cite{14DJ80}. Our strategy is somewhat different. We simply want to improve the description of the pseudoscalar mesons and we can do that by adjusting the values of the parameters $A$ and $R_{0}$ that govern the behaviour of the running coupling constant~(\ref{eq2.25}). The free parameters now are $B$, $\gamma$, $A$, $R_{0}$, and $m_{s}$, $m_{c}$, $m_{b}$. First, let us try to use the masses of four light hadrons (i.e. $N$, $\Delta$, $\rho$--$\omega$ system, and $\pi$) to fix the parameters $B$, $\gamma$, $A$, and $R_{0}$. Then, we employ $\phi$, $J/\psi$, and $\Upsilon$ to determine the masses of the strange, charmed, and bottom quarks. Predictions for the hadron mass values generated by this version of the model are presented in the column denoted as Mod3 of Tables~\ref{t3.1}--\ref{t3.3}. For the light hadrons now almost everything is all right. The description of the pseudoscalar mesons is improved considerably. The fit for the baryons is slightly worsened, but still remains of the quality similar to the original MIT version. The analysis of entries presented in Tables~\ref{t3.2} and \ref{t3.3} shows that in the heavy quark sector the meson spectrum is improved. However, for the baryons containing charmed quarks the discrepancy becomes more serious. A more careful analysis shows that the situation may be not so bad as appears. The hadron mass differences in the new version are described better, while the absolute position of the baryon spectrum is evidently positioned too low. Such regularities in the hadron spectrum is a welcomed feature, and we can conclude that this variant of the model can serve as a good starting point for a further development.
%TCIMACRO{\TeXButton{B}{\begin{table}[tbp] \centering}}
%BeginExpansion
\begin{table}[tbp] \centering%t05
%EndExpansion
\caption{Parameters for the six variants of the bag model as described in the text. All mass parameters ($m$, $\widetilde{m}$, $\delta $) are in GeV, $R_{0}$ in GeV$^{-1}$, $B$ in GeV$^{4}$.\label{t3.4}}
\begin{tabular}{lllllll}
\hline
Parameter & MIT & Mod1 & Mod2 & Mod3 & Mod4 & Mod5 \\ \hline
$B\cdot 10^{4}$ & 4.476 & 4.892 & 7.288 & 7.597 & 7.597 & 7.810 \\ 
$Z_{0}$ & -1.836 & -- & -- & -- & -- & -- \\ 
$\gamma $ & -- & 2.480 & 1.901 & 1.958 & 1.958 & 2.004 \\ 
$\alpha _{c}$ & 2.186 & 1.564 & 1.369 & -- & -- & -- \\ 
$A$ & -- & -- & -- & 1.070 & 1.070 & 0.622 \\ 
$R_{0}$ & -- & -- & -- & 2.543 & 2.543 & 4.473 \\ 
$m_{s}$ & 0.279 & 0.296 & 0.347 & 0.339 & -- & -- \\ 
$\widetilde{m}_{s}$ & -- & -- & -- & -- & 0.161 & 0.234 \\ 
$\delta _{s}$ & -- & -- & -- & -- & 0.156 & 0.101 \\ 
$m_{c}$ & 1.552 & 1.474 & 1.614 & 1.578 & -- & -- \\ 
$\widetilde{m}_{c}$ & -- & -- & -- & -- & 1.462 & 1.473 \\ 
$\delta _{c}$ & -- & -- & -- & -- & 0.109 & 0.095 \\ 
$m_{b}$ & 4.954 & 4.696 & 4.967 & 4.848 & -- & -- \\ 
$\widetilde{m}_{b}$ & -- & -- & -- & -- & 4.786 & 4.752 \\ 
$\delta _{b}$ & -- & -- & -- & -- & 0.069 & 0.089 \\ \hline
\end{tabular}
%TCIMACRO{\TeXButton{E}{\end{table}}}
%BeginExpansion
\end{table}%
%EndExpansion

Now we must find a way to improve the description of baryons and not to spoil the meson spectrum. Our proposal is to use for this purpose the scale-dependent effective quark mass given by the mass function~(\ref{eq2.26}). Instead of a fixed quark mass $m_{f}$ now we have two adjustable parameters $\widetilde{m}_{f}$ and $\delta_{f}$ for each quark flavour. For fixing these parameters we employ the masses of corresponding vector mesons ($\phi $, $J/\psi $, $\Upsilon $) and the mass values of the lightest baryons $\Lambda_{f}$ containing the quark $q_{f}$. The results of the fitting are given in the column denoted as Mod4 in Table~\ref{t3.4}. From Tables~\ref{t3.1}--\ref{t3.3} we see that the agreement with experiment is improved impressively. Despite the manifest success in describing the hadron mass spectrum, several imperfections of the model still remain uncured. One drawback common to almost all variants of the bag model is the $\Sigma$--$\Lambda$ mass difference. For the light hadrons it differs from the experimental value by about 30 MeV, and for the charmed hadrons the discrepancy of $\Sigma_{c}$--$\Lambda_{c}$ mass splitting from the experiment grows up to 60~MeV. One possible solution to this problem can be the use of some chiral extension of the bag model \cite{50MBX81,51S84,23RT85}, however, such an extension is outside the scope of the present investigation.

The other problem is the high sensitivity of the mass values of the pseudo\-scalar mesons to the changes of the parameters $A$ and $R_{0}$ that define the behaviour of the effective coupling constant~(\ref{eq2.25}). To illustrate the sensitivity of the model predictions upon the choice of the parameters $A$ and $R_{0}$ we present the results of an alternative calculation with somewhat different fitting procedure. The model parameters can be refitted to reproduce the kaon mass instead of the pion one. The corresponding results for the model parameters are presented in the column denoted as Mod5 of Table~\ref{t3.4}, and the results of calculations are given in the last column of Tables~\ref{t3.1}--\ref{t3.3}. We see that the model is rather stable in its predictions. Both versions (Mod4 and Mod5) provide a reasonable description of the ground state hadron spectrum. The most pronounced difference in the predicted hadronic mass values between the two versions is for the pseudoscalar mesons. It is impossible to fit the masses of the kaon and pion with a common set of parameters, and it teaches us that we cannot get all. Because the agreement with experiment is slightly better for the version denoted as Mod4 (with the fitted pion mass), we prefer to use this version of the model. However, our choice should not be taken too seriously. If one needs the model with the accurate kaon mass value one can use the version denoted as Mod5 as well.

There is also some concern about the masses of the heavy scalars (especially $\eta _{b}$). Light scalar mesons ($\eta $ and $\eta ^{\prime }$) need special treatment \cite{10DJJK75}, therefore, they are not included in our consideration. For the attempts to solve this problem by incorporating higher-order ($\sim \alpha _{c}^{2}$) corrections see \cite{52DG83,53SV88}. Meanwhile, it is a common practice to treat the heavy scalars on the same footing as the other hadrons. For the $\eta _{c}$ meson our prediction is $M(\eta _{c})=3.005$~GeV, which is about 25~MeV too high. This is an indication that the interaction strength for this state may be underestimated. One possible reason for this can be slightly too large value of the cavity radius $R(\eta _{c})$ that is obtained after the minimization of the hadron energy. The version of the model with the coupling constant and quark mass fixed (Mod2) gives $M(\eta _{c})=2.964$~GeV that is 15~MeV too low. By analogy we expect $M(\eta _{b})$ to lie somewhere in the region between 9.37 and 9.44~GeV (the corresponding potential model prediction is 9.40~GeV \cite{45CI86}). All these results are in some conflict with the recent experimental data $M_{\mathrm{ex}}(\eta _{b})=(9.30\pm 0.04)$~GeV \cite{54ALEPH02,42PDG02}. This result still needs additional confirmation. However, if confirmed, it could become a serious headache for the model builders. While the mass differences $M(B^{*})-M(B)$ and $M(B_{s}^{*})-M(B_{s})$ are reproduced with good accuracy, our result for the $\eta _{b}$ mass value $M(\eta _{b})=9.44$~GeV is evidently too high, signalling that scalars must be treated with care and in this particular case a more subtle treatment might be necessary.

In order to illustrate the main features of the model, some parameters characterizing the model (for the version Mod4) are given in Tables~\ref{t3.5}--\ref{t3.7}. By inspecting the entries presented in these tables one can see how the scale-dependent characteristics (coupling constant, mass values of the \textit{s-, c-, b}-quarks) change when going from one particle to another. For example, when going from the $\Delta $ baryon to $\eta _{b}$ meson, the strong coupling constant reduces by about 30\% from its maximum value $\alpha _{\max }=1.531$ to the minimum value $\alpha _{\min }=1.046$. The changes in the mass values are not so impressive and do not exceed 40~MeV for the strange, 30~MeV~-- for the charmed, and $\sim 25$~MeV~-- for the bottom quarks, respectively.
%TCIMACRO{\TeXButton{B}{\begin{table}[tbp] \centering}}
%BeginExpansion
\begin{table}[tbp] \centering%t06
%EndExpansion
\caption{Some characteristics of the bag model  (version Mod4) for hadrons consisting of the \textit{u-}, \textit{d-}, and \textit{s-}quarks. All masses and $E$ are given in GeV, $R$ in GeV$^{-1}$.\label{t3.5}}
\begin{tabular}{lllllll}
\hline
Particle & $M$ & $E$ & $R$ & $\beta (M^{2}/P^{2})$ & $\alpha _{c}(R)$ & $\overline{m}_{s}(R)$ \\ \hline
$N$ & 0.939 & 1.345 & 4.948 & 0.925 & 1.517 & -- \\ 
$\Delta $ & 1.232 & 1.561 & 5.015 & 0.944 & 1.531 & -- \\ 
$\pi $ & 0.137 & 0.857 & 4.417 & 0.854 & 1.401 & -- \\ 
$\rho ,\omega $ & 0.769 & 1.155 & 4.532 & 0.920 & 1.426 & -- \\ 
$\Lambda $ & 1.116 & 1.559 & 4.797 & 0.929 & 1.484 & 0.393 \\ 
$\Sigma $ & 1.159 & 1.596 & 4.757 & 0.931 & 1.476 & 0.392 \\ 
$\Xi $ & 1.310 & 1.784 & 4.639 & 0.932 & 1.450 & 0.388 \\ 
$\Sigma ^{*}$ & 1.388 & 1.758 & 4.883 & 0.944 & 1.503 & 0.396 \\ 
$\Xi ^{*}$ & 1.543 & 1.950 & 4.762 & 0.944 & 1.477 & 0.392 \\ 
$\Omega ^{-}$ & 1.695 & 2.135 & 4.652 & 0.945 & 1.452 & 0.388 \\ 
$K$ & 0.437 & 1.105 & 4.189 & 0.877 & 1.350 & 0.372 \\ 
$K^{*}$ & 0.896 & 1.340 & 4.393 & 0.920 & 1.395 & 0.379 \\ 
$\phi $ & 1.019 & 1.517 & 4.271 & 0.920 & 1.368 & 0.375 \\ \hline
\end{tabular}
%TCIMACRO{\TeXButton{E}{\end{table}}}
%BeginExpansion
\end{table}%
%EndExpansion

By comparing the values of $M$ and $E$ one can estimate the role and size of the c.m.m. correction for each particle. The typical correction is $\sim 400$~MeV for the light hadrons, $\sim 300$~MeV for the hadrons with charm, and $<250$~MeV for the hadrons containing bottom quarks.

Another interesting characteristic is the function $\beta (x)$ entering Eq.~(\ref{eq2.36}). From Table~\ref{t3.7} we see that for the bottom quarks $\beta (x)\approx 0.99$. In this case, in order to obtain the mass of the hadron one can use a simpler relation (\ref{eq2.38}) instead of Eq.~(\ref{eq2.35}). The difference between the two values calculated using Eq.~(\ref{eq2.35}) and Eq.~(\ref{eq2.38}) correspondingly will not exceed 1~MeV in this case. For the hadrons with charmed quarks (Table~\ref{t3.6}) the difference can grow up to $\approx 10$~MeV. This is not a very large difference, and the use of Eq.~(\ref{eq2.38}) to simplify calculations sometimes can be justifiable in this case, too. So we see that there remains only the light hadron sector where the results obtained by Eqs.~(\ref{eq2.35}) and (\ref{eq2.38}) may differ significantly.
%TCIMACRO{\TeXButton{B}{\begin{table}[tbp] \centering}}
%BeginExpansion
\begin{table}[tbp] \centering%t07
%EndExpansion
\caption{Some characteristics of the bag model  (version Mod4) for hadrons with charmed quarks. All masses and $E$ are given in GeV, $R$ in GeV$^{-1}$.\label{t3.6}}
\begin{tabular}{llllllll}
\hline
Particle & $M$ & $E$ & $R$ & $\beta (M^{2}/P^{2})$ & $\alpha _{c}(R)$ & $\overline{m}_{s}(R)$ & $\overline{m}_{c}(R)$ \\ \hline
$\Lambda _{c}^{+}$ & 2.285 & 2.597 & 4.588 & 0.965 & 1.439 & -- & 1.618 \\ 
$\Sigma _{c}$ & 2.392 & 2.695 & 4.572 & 0.967 & 1.435 & -- & 1.618 \\ 
$\Xi _{c}$ & 2.466 & 2.817 & 4.428 & 0.964 & 1.403 & 0.380 & 1.615 \\ 
$\Xi _{c}^{\prime }$ & 2.546 & 2.885 & 4.449 & 0.966 & 1.408 & 0.381 & 1.615 \\ 
$\Omega _{c}^{0}$ & 2.696 & 3.070 & 4.336 & 0.965 & 1.383 & 0.377 & 1.612 \\ 
$\Xi _{cc}$ & 3.556 & 3.865 & 4.205 & 0.976 & 1.353 & -- & 1.609 \\ 
$\Omega _{cc}^{+}$ & 3.709 & 4.051 & 4.094 & 0.975 & 1.328 & 0.369 & 1.606 \\ 
$\Sigma _{c}^{*}$ & 2.488 & 2.770 & 4.668 & 0.970 & 1.456 & -- & 1.620 \\ 
$\Xi _{c}^{*}$ & 2.637 & 2.954 & 4.547 & 0.969 & 1.429 & 0.384 & 1.617 \\ 
$\Omega _{c}^{0*}$ & 2.782 & 3.133 & 4.437 & 0.967 & 1.405 & 0.381 & 1.615 \\
$\Xi _{cc}^{*}$ & 3.659 & 3.948 & 4.311 & 0.978 & 1.377 & -- & 1.612 \\ 
$\Omega _{cc}^{+*}$ & 3.799 & 4.119 & 4.204 & 0.976 & 1.353 & 0.372 & 1.609 \\ 
$\Omega _{ccc}^{++}$ & 4.776 & 5.092 & 3.954 & 0.981 & 1.296 & -- & 1.603 \\ 
$D$ & 1.833 & 2.210 & 3.934 & 0.953 & 1.292 & -- & 1.602 \\ 
$D^{*}$ & 2.002 & 2.327 & 4.115 & 0.960 & 1.333 & -- & 1.607 \\ 
$D_{s}$ & 1.965 & 2.335 & 3.807 & 0.951 & 1.262 & 0.358 & 1.599 \\ 
$D_{s}^{*}$ & 2.119 & 2.494 & 3.995 & 0.958 & 1.306 & 0.365 & 1.604 \\ 
$\eta _{c}$ & 3.005 & 3.397 & 3.545 & 0.967 & 1.202 & -- & 1.593 \\ 
$J/\psi $ & 3.097 & 3.454 & 3.689 & 0.970 & 1.235 & -- & 1.596 \\ \hline
\end{tabular}
%TCIMACRO{\TeXButton{E}{\end{table}}}
%BeginExpansion
\end{table}%
%EndExpansion

We want to end up this section with several comments. The attentive reader could already have noticed (look at the values of $\gamma$ in Table~\ref{t3.4}) that our effective momentum square (\ref{eq2.31}) is about twice the value usually accepted in the various bag model calculations (see e.g. \cite{16LW82,17CHP83,22BSRMT84,23RT85} etc.). It must be so, and we'll soon see why.
%TCIMACRO{\TeXButton{B}{\begin{table}[tbp] \centering}}
%BeginExpansion
\begin{table}[tbp] \centering%t08
%EndExpansion
\caption{Some characteristics of the bag model (version Mod4) for the lightest hadrons with bottom quarks. All masses and $E$ are given in GeV, $R$ in GeV$^{-1}$.\label{t3.7}}
\begin{tabular}{lllllllll}
\hline
Particle & $M$ & $E$ & $R$ & $\beta (M^{2}/P^{2})$ & $\alpha _{c}(R)$ & $\overline{m}_{s}(R)$ & $\overline{m}_{c}(R)$ & $\overline{m}_{b}(R)$ \\ 
\hline
$\Lambda _{b}^{0}$ & 5.624 & 5.777 & 4.453 & 0.991 & 1.409 & -- & -- & 4.883 \\ 
$\Sigma _{b}$ & 5.759 & 5.906 & 4.478 & 0.992 & 1.414 & -- & -- & 4.883 \\ 
$B$ & 5.252 & 5.425 & 3.774 & 0.990 & 1.255 & -- & -- & 4.873 \\ 
$B^{*}$ & 5.309 & 5.472 & 3.870 & 0.990 & 1.277 & -- & -- & 4.874 \\ 
$B_{s}$ & 5.387 & 5.591 & 3.642 & 0.988 & 1.224 & 0.352 & -- & 4.870 \\ 
$B_{s}^{*}$ & 5.439 & 5.632 & 3.741 & 0.989 & 1.247 & 0.356 & -- & 4.872 \\ 
$B_{c}$ & 6.307 & 6.550 & 3.297 & 0.988 & 1.143 & -- & 1.586 & 4.865 \\ 
$B_{c}^{*}$ & 6.345 & 6.576 & 3.378 & 0.989 & 1.162 & -- & 1.588 & 4.866 \\ 
$\eta _{b}$ & 9.438 & 9.661 & 2.894 & 0.992 & 1.046 & -- & -- & 4.858 \\ 
$\Upsilon $ & 9.460 & 9.675 & 2.950 & 0.993 & 1.060 & -- & -- & 4.859 \\ 
\hline
\end{tabular}
%TCIMACRO{\TeXButton{E}{\end{table}}}
%BeginExpansion
\end{table}%
%EndExpansion

Let us compare our method to deal with the c.m.m. with the methods used in Refs.~\cite{16LW82} and \cite{22BSRMT84,23RT85}. We expect our approach to give similar results for the light hadrons as the others do, because the masses of the light hadrons are used as an input to determine the basic model parameters. In the approach advocated in \cite{16LW82} the energy of the hadron is divided into two parts: $E_{\mathrm{q}}\sim 1/R$ associated with the quarks, and the volume energy
\begin{equation}
E_{V}=\frac{4\pi }{3}BR^{3},  \label{eq3.04}
\end{equation}
associated with a bag. Only the part $E_{\mathrm{q}}$ associated with the quarks is corrected for the c.m.m.
\begin{equation}
E_{\mathrm{q}}^{\mathrm{cor}}=\left( E_{\mathrm{q}}^{2}-P_{0}^{2}\right) ^{\frac{1}{2}},  \label{eq3.05}
\end{equation}
where $P_{0}^{2}$ is given by the analogue of Eq.~(\ref{eq2.31}) with $\gamma_{0}=1$. The total c.m.m.-corrected energy now can be written as
\begin{equation}
E^{\mathrm{cor}}=E_{\mathrm{q}}^{\mathrm{cor}}+E_{V},  \label{eq3.06}
\end{equation}
and the minimum of this energy is assumed to be the actual hadron mass. If the uncorrected energy 
\begin{equation}
E=E_{\mathrm{q}}+E_{V}  \label{eq3.07}
\end{equation}
is minimized first (as in Refs.~\cite{22BSRMT84,23RT85} and in our work), then the spurious energy of the center-of-mass motion is confined inside the bag, too. From the dimensional analysis it follows that for the massless quarks 
\begin{equation}
E=\frac{4}{3}E_{\mathrm{q}},  \label{eq3.08}
\end{equation}
and therefore, in order to subtract the spurious c.m.m. energy, one is forced to employ the relation of the type~(\ref{eq2.38}) in which the effective momentum square with
\begin{equation}
\gamma \geq \left( \frac{4}{3}\right) ^{2}  \label{eq3.09}
\end{equation}
must be used. This is an exact result (i.e. $\gamma =(4/3)^{2}$) in a toy model with massless noninteracting quarks for the particular hadron chosen in the fitting procedure while determining the bag constant. In our work we treat the parameter $\gamma$ as a quantity to be fitted. The values obtained (see Table~\ref{t3.4}) favour well the Eq.~(\ref{eq3.09}). For the version Mod1 the value of this parameter is substantially larger because in this case we must also subtract the ``spurious'' self-energy. So we see that because of its semi-phenomenological nature the effective momentum square~(\ref{eq2.31}) cannot be associated with a pure c.m.m. correction, but it may also contain some other $\sim 1/R$ corrections.

We have established the link between the treatment of Ref.~\cite{16LW82} and ours. Now it is evident that the version Mod1 of our treatment is similar to the model used in \cite{16LW82} and, therefore, in the case of the light hadrons, both models should give qualitatively similar results. The prescription advocated in Ref.~\cite{16LW82} seems to be somewhat more physical. However, this approach is hardly compatible with the Eq.~(\ref{eq2.35}) we have used to employ the c.m.m. correction.

The approach adopted by authors of Refs.~\cite{22BSRMT84,23RT85} is more similar to ours. They minimized the ``c.m.m.-uncorrected'' energy with the self-energy term included, and used Eq.~(\ref{eq2.38}) to incorporate the c.m.m. correction. However, their ``uncorrected'' energy contains the so-called zero-point energy term, which can be interpreted as representing mostly a c.m.m. correction \cite{16LW82}. Their zero-point constant $Z_{0}\approx -0.8$ is much smaller in comparison with the original MIT value $Z_{0}\approx -1.84$. So, the expression of the energy adopted in Refs.~\cite{22BSRMT84,23RT85} can be considered to be partially c.m.m.-corrected. In other words, in their approach the c.m.m. correction is incorporated in two steps. First, the term $Z_{0}/R$ is subtracted from the energy and this partially corrected energy is minimized. Then, Eq.~(\ref{eq2.38}) is applied, in which the usual expression for the momentum square of the quarks confined in the bag, $P_{0}^{2}=\sum n_{i}p_{i}^{2}$, is used. Eventually, one obtains similar results as in Ref. ~\cite{16LW82}. At present we have no simple answer to the question what quantity, the energy or the mass (c.m.m.-corrected energy), must be minimized (see the discussion on this subject in Ref.~\cite{19D81}). As we have seen, in practice this is somewhat a matter of taste, and seemingly rather different methods may give quite similar results.

\section{Electroweak properties}

The bag model also sets a framework to calculate other static properties of the hadrons. In this section we present our results for some electroweak properties: magnetic moments, charge radii, and axial-vector coupling constant. To obtain some feeling for the sensitivity of computed quantities to the model assumptions, we compare our predictions with the results of the original MIT model and with the experiment. Magnetic moments of the hadrons in the static spherical cavity approximation can be represented in the form \cite{09T84} (see also \cite{32ERR85}):
\begin{equation}
\mu _{\mathrm{h}}^{0}=\sum\limits_{i}\mu _{\mathrm{h}}^{i}\left\langle \mathrm{h}\right| \mathbf{\sigma 
}_{z}^{i}q_{i}\left| \mathrm{h}\right\rangle ,  \label{eq4.01}
\end{equation}
where $q_{i}$ is the charge of the \textit{i}-th quark and parameters $\mu^{i}$ are given by \cite{10DJJK75}:
\begin{equation}
\mu ^{i}=\frac{4\varepsilon _{i}R_{\mathrm{h}}+2m_{i}R_{\mathrm{h}}-3}{2(\varepsilon_{i}R_{\mathrm{h}}-1)\varepsilon _{i}R_{\mathrm{h}}+m_{i}R_{\mathrm{h}}}\ \frac{R_{\mathrm{h}}}{6}.
\label{eq4.02}
\end{equation}
In the last expression $\varepsilon _{i}$ represents the energy of the \textit{i}-th quark and $R_{\mathrm{h}}$ stands for the bag radius of the hadron under consideration. Magnetic transition moments are defined by
\begin{equation}
\mu _{\mathrm{h}\rightarrow \mathrm{h}^{\prime }}^{0}=\left\langle \mathrm{h}^{\prime }\right| \mathbf{\sigma }_{z}^{i}q_{i}\left| \mathrm{h}\right\rangle ,  \label{eq4.03}
\end{equation}
where $R_{\mathrm{h}}=R_{\mathrm{h}^{\prime }}$ is assumed. 

Matrix elements $\left\langle \mathrm{h}^{\prime }\right| \mathbf{\sigma }_{z}^{i}q_{i}\left| \mathrm{h}\right\rangle $ can be calculated with SU(6) wave functions as described in Ref.~\cite{55C79}, and for the cases we are interested in are displayed in Table~\ref{t4.1}.
%TCIMACRO{\TeXButton{B}{\begin{table}[tbp] \centering} }
%BeginExpansion
\begin{table}[tbp] \centering%t09
%EndExpansion
\caption{Composition of baryon magnetic moments. The label $l$ is used to collectively represent up and down quarks.\label{t4.1}}
\begin{tabular}{|c|c|}
\hline
Particle h & $\mu _{\mathrm{h}}^{0}$ \\ \hline
$P$ & $\mu ^{l}$ \\ \hline
$N$ & $-\frac{2}{3}\mu ^{l}$ \\ \hline
$\Sigma ^{+}$ & $\frac{1}{9}(8\mu ^{l}+\mu ^{s})$ \\ \hline
$\Sigma ^{0}$ & $\frac{1}{9}(2\mu ^{l}+\mu ^{s})$ \\ \hline
$\Sigma ^{-}$ & $\frac{1}{9}(\mu ^{s}-4\mu ^{l})$ \\ \hline
$\Lambda $ & $-\frac{1}{3}\mu ^{s}$ \\ \hline
$\Xi ^{0}$ & $-\frac{2}{9}(\mu ^{l}+2\mu ^{s})$ \\ \hline
$\Xi ^{-}$ & $\frac{1}{9}(\mu ^{l}-4\mu ^{s})$ \\ \hline
$\Sigma ^{0}\rightarrow \Lambda $ & $-\frac{1}{\sqrt{3}}\mu ^{l}$ \\ \hline
$\Omega ^{-}$ & $-\mu ^{s}$ \\ \hline
$\Delta ^{++}$ & $2\mu ^{l}$ \\ \hline
\end{tabular}
%TCIMACRO{\TeXButton{E}{\end{table}}}
%BeginExpansion
\end{table}%
%EndExpansion

The square charge radius and axial-vector coupling constant can be calculated from the expressions
\begin{equation}
\left\langle r_{0}^{2}\right\rangle _{\mathrm{h}}=\sum\limits_{i}q_{i}\int\limits_{0}^{R_{\mathrm{h}}}r^{2}dr\left[P_{i}^{2}(r)+Q_{i}^{2}(r)\right] ,  \label{eq4.04}
\end{equation}

\begin{equation}
g_{A}^{0}=\frac{5}{3}\int\limits_{0}^{R_{\mathrm{h}}}dr\left[ P^{2}(r)-\frac{1}{3} Q^{2}(r)\right] .  \label{eq4.05}
\end{equation}
Analytic expressions for Eqs.~(\ref{eq4.04}) and (\ref{eq4.05}) can be found in Ref.~\cite{10DJJK75}. In the case of massless quarks the value of $g_{A}^{0}$ does not depend on the radius $R_{\mathrm{h}}$ and is always equal to 1.088. 

Before comparing the static quantities with the corresponding experimental values they must be corrected for the center-of-mass motion. For this purpose we adopt the prescription proposed in Refs.~\cite{24BSMT84,25S84} (see also \cite{32ERR85}). Their formulae for the corrected values of $\mu $, $\left\langle r^{2}\right\rangle $, and $g_{A}$ are \cite{24BSMT84}:
\begin{equation}
\mu _{\mathrm{h}}=\frac{3}{1+\left\langle M/E\right\rangle +\left\langle M^{2}/E^{2}\right\rangle }\left[ \mu _{\mathrm{h}}^{0}+\frac{1-\left\langle M/E\right\rangle }{3}\frac{M_{\mathrm{P}}}{M_{\mathrm{h}}}Q_{\mathrm{h}}\right] ,  \label{eq4.06}
\end{equation}

\begin{equation}
\left\langle r^{2}\right\rangle _{\mathrm{h}}=\frac{3}{1+2\left\langle M^{2}/E^{2}\right\rangle }\left[ \left\langle r_{0}^{2}\right\rangle _{\mathrm{h}}-\frac{9Q_{\mathrm{h}}}{4P^{2}}\right] ,  \label{eq4.07}
\end{equation}

\begin{equation}
g_{A}=\frac{3}{1+2\left\langle M/E\right\rangle }g_{A}^{0}.  \label{eq4.08}
\end{equation}
In the equations above, the values of $\left\langle M/E\right\rangle $ and $\left\langle M^{2}/E^{2}\right\rangle $ are defined by Eqs.~(\ref{eq2.33}) and (\ref{eq2.34}), respectively, $P^{2}$ is given by Eq.~(\ref{eq2.31}), $M_{\mathrm{P}}$ is the mass of the proton, $Q_{\mathrm{h}}$ and $M_{\mathrm{h}}$ stand for the charge and the mass of the corresponding hadron.

Our predictions for the magnetic moments are presented in Table~\ref{t4.2}. In order to simplify the calculation of the transition moment $\mu(\Sigma ^{0}\rightarrow \Lambda )$, the same wave function (that of the $\Sigma $ baryon) was used for both states. The experimental values were taken from the Particle Data Group \cite{42PDG02}.
%TCIMACRO{\TeXButton{B}{\begin{table}[tbp] \centering} }
%BeginExpansion
\begin{table}[tbp] \centering%t10
%EndExpansion
\caption{Magnetic moments of baryons (in nuclear magnetons). Uncorrected values are enclosed in parentheses.\label{t4.2}}
\begin{tabular}{ccccccc}
\hline
Particles  & $M_{\mathrm{ex}}$ & MIT & \multicolumn{2}{c}{Mod1} & \multicolumn{2}{c}{Mod4} \\ \hline
$P$ & 2.79 & 1.90 & (2.08) & 2.89 & (1.88) & 2.61 \\ 
$N$ & -1.91 & -1.26 & (-1.39) & -1.85 & (-1.25) & -1.66 \\ 
$\Sigma ^{+}$ & 2.46 & 1.83 & (2.01) & 2.67 & (1.75) & 2.34 \\ 
$\Sigma ^{0}$ & -- & 0.58 & (0.63) & 0.81 & (0.54) & 0.70 \\ 
$\Sigma ^{-}$ & -1.16 & -0.67 & (-0.75) & -1.05 & (-0.66) & -0.94 \\ 
$\Xi ^{0}$ & -1.25 & -1.05 & (-1.13) & -1.44 & (-0.95) & -1.22 \\ 
$\Xi ^{-}$ & -0.65 & -0.43 & (-0.45) & -0.64 & (-0.36) & -0.54 \\ 
$\Lambda $ & -0.61 & -0.48 & (-0.51) & -0.66 & (-0.43) & -0.55 \\ 
$\Sigma ^{0}\rightarrow \Lambda $ & -1.61 & -1.08 & (-1.19) & -1.54 & (-1.05) & -1.35 \\ 
$\Omega ^{-}$ & -2.02 & -1.54 & (-1.56) & -1.91 & (-1.26) & -1.57 \\ 
$\Delta ^{++}$ & 3.7$\div $7.5 & 4.16 & (4.36) & 5.40 & (3.81) & 4.76 \\ 
\hline
\end{tabular}
%TCIMACRO{\TeXButton{E}{\end{table}}}
%BeginExpansion
\end{table}%
%EndExpansion

Predictions for other electroweak properties are given in Table~\ref{t4.3}. The experimental values for $r_{\mathrm{P}}$, $r_{_{\Sigma ^{-}}}$, $r_{\pi }$, and $r_{K}$ were taken from Refs.~\cite{55R00}, \cite{56S00}, \cite{57DEA82}, and \cite{58AEA80}, respectively.
%TCIMACRO{\TeXButton{B}{\begin{table}[tbp] \centering} }
%BeginExpansion
\begin{table}[tbp] \centering%t11
%EndExpansion
\caption{Some electroweak parameters of the hadrons. All charge radii are given in fm. Uncorrected values are enclosed in parentheses.\label{t4.3}}
\begin{tabular}{ccccccc}
\hline
Parameter  & $M_{\mathrm{ex}}$ & MIT & \multicolumn{2}{c}{Mod1} & \multicolumn{2}{c}{Mod4} \\ \hline
$r_{\mathrm{P}}$ & 0.88 & 0.73 & (0.79) & 0.80 & (0.71) & 0.70 \\ 
$r_{_{\Sigma ^{-}}}$ & 0.77 & 0.68 & (0.76) & 0.77 & (0.66) & 0.66 \\ 
$r_{\pi }$ & 0.66 & 0.49 & -- & -- & (0.63) & 0.66 \\ 
$r_{K}$ & 0.58 & 0.47 & (0.65) & 0.70 & (0.58) & 0.60 \\ 
$g_{A}$ & 1.26 & 1.09 & (1.09) & 1.34 & (1.09) & 1.33 \\ \hline
\end{tabular}
%TCIMACRO{\TeXButton{E}{\end{table}}}
%BeginExpansion
\end{table}%
%EndExpansion

Predicted values in the variants Mod2--Mod5 of the model are of similar quality (differences between the calculated values do not exceed 5\%), therefore, we list only the values obtained using variants Mod1 and Mod4. Agreement with the experimental values is good in both variants of the model. Predictions for the electroweak parameters calculated in the variant Mod4 are comparable to the results obtained in Ref.~\cite{22BSRMT84}. 

Predicted values of the electroweak parameters practically are insensitive to the corrections associated with the scale dependence of the effective coupling constant and the quark mass. In contrast, the c.m.m. corrections for magnetic moments and for axial-vector coupling constant improve the predictions significantly. The corrections for the charge radii in all cases are of minor importance. Neutron charge radii in our version of the model remain zeroes. This drawback of the model is a direct consequence of the isospin symmetry. 

Our feeling is that one must not take the good agreement of the corrected values of the electroweak parameters with the experiment too seriously. For example, the version Mod1, in which the c.m.m. correction seems to be overestimated, better agrees with the experiment than the version Mod4. In fact, because we have employed the c.m.m. correction in a somewhat phenomenological fashion, we cannot disentangle the pure center-of-mass motion and other possible effects. The treatment of c.m.m. in our work is far from being perfect. Moreover, other effects, such as recoil corrections and the pion cloud contribution, may also be important. Nevertheless, despite of all this criticism, the model seems to provide reasonable predictions for the electroweak properties of the hadrons. 

\section{Summary}

One of the objectives of this paper was to examine the influence of the corrections associated with the center-of-mass motion, the scale dependence of the running coupling constant, and the scale dependence of effective quark mass on the mass spectrum and on other static properties of the hadrons, calculated in the framework of the bag model. Special attention is paid to the hadrons containing heavy (charmed and bottom) quarks. All quarks are treated on equal footing. The heavy quarks rattle inside the bag cavity in the manner of the light ones with the maximum of their distribution being closer to the center of the bag than the same of the light quarks.

Incorporating the corrections consecutively we were able to investigate the effect of these corrections upon the predictions of the model. We have found that the proper treatment of the center-of-mass motion is essential to obtain the reasonable description of the heavy hadrons. The running coupling constant and scale-dependent effective quark mass proved to be useful ingredients of the modified bag model. These corrections helped us to obtain the good agreement of the calculated masses of the heavy hadrons with the experimental values. There is strong evidence that these two modifications of the model should be applied simultaneously. 

The bag model with all these corrections included can be treated as rather simple and controllable framework for the unified description of the light and heavy hadrons. Maybe the worst discrepancy of the model is the $\pi $--$K$ mass difference. Another systematic discrepancy inherited from the original MIT version of the model is the $\Sigma _{\mathrm{h}}$--$\Lambda _{\mathrm{h}}$ mass splitting. Finally, the description of the $\eta _{\mathrm{h}}$ states also seems to be somewhat problematic. Despite the several drawbacks mentioned above, the model accounts reasonably well for the masses of almost all hadrons under investigation. The accuracy achieved in the description of the hadron spectrum suggests that for the further improvement an explicit breaking of the isospin symmetry may be necessary.

\end{document}